\documentclass{article}

\usepackage{arxiv}

\usepackage[utf8]{inputenc} % allow utf-8 input
\usepackage[T1]{fontenc}    % use 8-bit T1 fonts
\usepackage{hyperref}       % hyperlinks
\usepackage{url}            % simple URL typesetting
\usepackage{booktabs}       % professional-quality tables
\usepackage{amsfonts}       % blackboard math symbols
\usepackage{nicefrac}       % compact symbols for 1/2, etc.
\usepackage{microtype}      % microtypography
\usepackage{lipsum}
\usepackage{graphicx}
\usepackage{placeins}
\usepackage{soul,color}
\usepackage{makecell}
\graphicspath{ {./images/} }

\title{MaterialsAtlas.org: A Materials Informatics Web App Platform for Materials Discovery and Survey of State-of-the-Art}

\author{
 Jianjun Hu *\\
 Department of Computer Science and Engineering\\
  University of South Carolina\\
  Columbia, SC 29201 \\
  \texttt{jianjunh@cse.sc.edu} \\
    \And
 Stanislav Stefanov,  Yuqi Song, Sadman Sadeed Omee\\
 Department of Computer Science and Engineering\\
  University of South Carolina\\
  Columbia, SC 29201 \\
  \And
 Steph-Yves Louis, Edirisuriya M. D. Siriwardane, Yong Zhao \\
 Department of Computer Science and Engineering\\
  University of South Carolina\\
  Columbia, SC 29201 \\
}

\begin{document}
\maketitle
\begin{abstract}

The availability and easy access of large scale experimental and computational materials data have enabled the emergence of accelerated development of algorithms and models for materials property prediction, structure prediction, and generative design of materials. However, lack of user-friendly materials informatics web servers has severely constrained the wide adoption of such tools in the daily practice of materials screening, tinkering, and design space exploration by materials scientists. Herein we first survey current materials informatics web apps and then propose and develop MaterialsAtlas.org, a web based materials informatics toolbox for materials discovery, which includes a variety of routinely needed tools for exploratory materials discovery, including materials composition and structure check (e.g. for neutrality, electronegativity balance, dynamic stability, Pauling rules), materials property prediction (e.g. band gap, elastic moduli, hardness, thermal conductivity), and search for hypothetical materials. These user-friendly tools can be freely accessed at \url{www.materialsatlas.org}. We argue that such materials informatics apps should be widely developed by the community to speed up the materials discovery processes.

\end{abstract}

% keywords can be removed

\section{Introduction}

Machine learning(ML) models and algorithms are increasingly applied in materials science for a wide variety of tasks ranging from materials characterization, property prediction, and to structure/composition generation design as reviewed in  \cite{raccuglia2016machine,ramprasad2017machine,gubernatis2018machine,butler2018machine,wei2019machine,morgan2020opportunities,wang2020machine,chen2020machine,moosavi2020role,saal2020machine,sparks2020machine}. These data driven algorithms have dramatically sped up the exploration in the vast chemical design space and have helped to discover many novel functional materials\cite{chen2020critical}. However, compared to the mature bioinformatics field with thousands of web servers (>9000) \cite{fehlmann2021aviator,kern2020lifetime}, the ecosystem of materials informatics is still in the embryo stage with less than 100 web servers with most of them being data infrastructures \cite{himanen2019data}. This can also be seen in our survey in Table 1 which focuses on inorganic crystal materials. In contrast, the bioinformatics field even has a search engine bio.tools which indexes and tracks biological scientific web servers along their lifetime. 

Here we argue that despite the increased sharing of data, programs or source code in the materials informatics community, the missing web apps for these tools have significantly impeded the progress of our field as most experimentalist teams do not have the expertise to implement, train and deploy these tools locally and many of the proposed materials informatics algorithms are under-used. Indeed, compared to bioinformatics, materials informatics web tools are much fewer in terms of quantity, diversity and quality. Developing and providing web servers can make complex algorithms accessible to a broad research and user community. In addition to providing user-friendly services to materials researchers, a recent study has found that there exists a positive association between the number of citations and the probability of a web server being reachable \cite{kern2020lifetime}. 

Currently, the most widely used web services in materials include Materials Project(MP) \cite{ceder2010materials}, Aflow-lib\cite{curtarolo2012aflowlib}, and OQMD \cite{kirklin2015open}, which are all mainly used as data sources. Even though these major databases come with several related analysis tools, there are many missing web apps that are strongly needed in an exploratory materials discovery process. This process can be generally divided into four major stages each needing specific convenient web apps: characterization, property prediction, synthesis, theory discovery, and materials design \cite{li2020ai}. 

Starting from the composition exploration, one would need tools and models that can check the charge neutrality and electronegativity balance and estimate its formation energy. Composition based prediction of crystal symmetry or lattice constants or even crystal structures are also highly desirable. When structures can be predicted or obtained via element substitution, tools such as structural relaxation, formation energy calculation, e-above-hull energy calculation, Pauling rule check, phonon calculation, and synthesizability are all useful to evaluate the feasibility of the candidate materials. The second major category of tools needed are property prediction web apps as provided by several existing servers \cite{curtarolo2012aflowlib, choudhary2020joint}. However, many of these property prediction web apps do not support screening multiple inputs, which limit their usage in high-throughput screening for new materials. Nowadays, the modern deep generative materials design models can easily generate millions of candidate compositions \cite{dan2020generative} and structures \cite{zhao2021high}. Also, many of these tools also do not support convenient download of the prediction results. In addition, it is desirable that databases of hypothetical new materials can be made available for users to find novel functional materials. 

In this paper, we first survey current state-of-the-art web services in the inorganic materials community and identify the requirements of a good materials web app and the limitations of current web apps. We then introduce MaterialsAtlas.org, our materials informatics web app platform for supporting the whole life cycle of materials discovery. It includes multiple candidate materials composition and structure validations/checks, materials property prediction modules, hypothetical materials databases, and utility tools. Our web apps are developed with high-throughput materials discovery processes in mind with a user-friendly web interface and an easy download of results.

\section{Survey of existing web apps for materials discovery}

 While there are many known AI or ML studies applied to the materials discovery process \cite{saal2020machine,lu2021computational}, many of them do not offer or share their code, programs, web apps, or even datasets, which significantly lower their potentials in materials research. Compared to thousands of bioinformatics web apps, the number of materials informatics web apps are much fewer and are developed in an ad hoc way without considering the high-throughput screening requirement from the materials discovery process. Table\ref{tab:webapp} shows a list of web apps and tools that support the materials discovery process. 

Materials characterization is a key step in experimental analysis which is especially true with the progress of high-throughput materials characterization that generates huge amounts of data. There are an increasing number of algorithmic studies on phase mapping of X-ray diffraction data \cite{stanev2018unsupervised,xiong2017automated},  symmetry determination in electron diffraction\cite{kaufmann2020crystal},  predicting crystallographic dimensionality and space group from a limited number of thin-film XRD patterns \cite{oviedo2019fast}, predicting accurate scale factor, lattice parameter and crystallite size maps for all phases \cite{dong2021deep}, and tuning of parameters in the Rietveld method
\cite{ozaki2020automated}. However, most of these studies provide user-friendly web services. In our survey, only USCD team provides a web tool for coordination environment prediction from X-ray absorption spectroscopy\cite{zheng2020random}. 
 
The second major category of web tools is for materials property prediction. This includes aflow-ML\cite{curtarolo2012aflowlib}, Javis-ML\cite{choudhary2020joint}, Crystal.AI\cite{crystalsai}, thermoelectric predictor \cite{gaultois2016perspective}, NIMS tools \cite{tanifuji2019materials}, SUNCAT catalysis property predictor \cite{suncat}, and matlearn\cite{peterson2021materials}. These web apps cover a variety of materials properties. For example, JARVIS-ML from NIST can predict formation energies, exfoliation energies, bandgaps, magnetic moments, refractive index, dielectric, thermoelectric, and maximum piezoelectric and infrared modes. However, many of these web apps are developed in an ad hoc way; they usually only accept one composition or structure at a time and cannot be used for screening. They usually do not come up with a performance measure to indicate the prediction confidence. More importantly, many of the algorithms or descriptors are outdated. For example a recent benchmark studies showed that the best algorithms for formation energy and bandgap prediction are based on graph neural networks, which are all much better than other structural descriptor based methods as used in \cite{curtarolo2012aflowlib} and \cite{choudhary2020joint}.

% \usepackage[normalem]{ulem}
% \useunder{ine}{}{}
\begin{table}[ht!]
% \resizebox{\textwidth}{!}{% use resizebox with textwidth
\centering
\caption{Survey of current web apps for materials discovery}
\label{tab:webapp}
% \begin{tabular}{|l|l|l|l|l|}
\begin{tabular}{|p{2cm}|p{3.6cm}|p{1.7cm}|p{6cm}|p{2cm}|}

\hline
App Name                       & URL                                                                             & Institute                                                                       & App functions                                                                                                                                                                                                         & Comment                                                                                                                   \\ \hline
MaterialsAtlas                 &  \url{ www.materialsatlas.org}                      & UofSC                                                                           & composition/structure validation, property prediction, screening of materials, ML, composition enumeration and more                                                                                      & this work. easy to use.                                                                                                                          \\ \hline
Materials Project\cite{ceder2010materials}               & { \url{ materialsproject.org}}                      & Lawrence Berkeley National Lab                                           & crystal toolkit, structure predictor, phase diagram, Pourbaix Diagram, reaction calculator, interface reaction, nanoporous materials analysis, synthesis description search                             &  major public repository. good web apps.                                                                                                                         \\ \hline
Aflowlib\cite{curtarolo2012aflowlib}                          & { \url{ aflowlib.org}}                               & Duke                                                                            & elastic, thermal, prototype, chull, aflow-ML for superconductor Tc, free energy and entropy, metal/insulator classification, band gap energy, bulk/shear moduli, Debye temperature and heat capacities                & outdated descriptor methods                                                                                              \\ \hline
OQMD\cite{kirklin2015open}                           & { \url{ oqmd.org/analysis}}           &   \makecell*[l]{North\\western}                                                                    & phase diagram, structure visualizer, ground state analysis                                                                                                                                                            & limited analysis web apps                                                                                                                          \\ \hline
JARVIS\cite{choudhary2020joint}                          & { \url{ jarvis.nist.gov}}                           & NIST                                                                            & web ML tools for diverse property predictions (regression/classifications)                                                                                                                                                   & account needs approval. CFID descriptors.                                                                                                                          \\ \hline
Crystal.AI \cite{zheng2020random}                 & { \url{ crystals.ai}}                           & UCSD                                                                           & prediction models of formation energy, bandgap, elastic constants, perovskite/garnet stability, Coordination from X-ray absorption spectroscopy & characterization and property prediction                                       \\ \hline
% Matgenb                        & { { matgenb.materialsvirtuallab.org}}            & UCSD                                                                            & code examples for materials informations problems                                                                                                                                                                     & not web app.                                                                                                                        \\ \hline

Matgenie\cite{matgenie}                       &   \url{matgenie.materialsvirtuallab.org}    & UCSD                                                                            & {materials analysis web app. structure file format conversion;symmetry analysis; structure  similarity comparison; XRD calculation; surface generation}                                                                                                                                                       & utility tools \\ \hline

Materials Cloud\cite{talirz2020materials}                 &  \url{materialscloud.org/work/tools}  & EPFL                                                                            & QE input generator, chemical shift, molecular polarizability, phonon visualizer, synthesis condiction finder, predicting oxidation states, atomic environment finder, electron transport, simulation in cloud (AiiDA) & mainly utility tools                                                                                                                          \\ \hline

Bilbao crystallographic server\cite{aroyo2006bilbao} & { \url{ www.cryst.ehu.es}}                          & {Univ.of Basque Country} & show Wyckoff positions, symmetry, structure utility                                                                                                                                                                   &   utility tools                                                                                                                        \\ \hline
\makecell*[l]{Thermoelectric\\\cite{gaultois2016perspective} }                & \url{thermoelectrics.citrination.com}        & Citrine                                                                         & predict thermoelectric materials properties                                                                                                                                                                           &  commercial solution                                                                                                                        \\ \hline

NIMS\cite{tanifuji2019materials} & \url{mits.nims.go.jp/en/} & Japan Nat. inst. of Mat. Sci. & Various databases and Composite Design \& Property Prediction System & rich databases \&Data Conversion Tools
\\ \hline
SUNCAT\cite{winther2019catalysis} & \url{catalysis-hub.org} & Stanford Univ. & database and tools for interface science and catalysis design & diagrams, ML models, diverse tools \\ 
\hline 

Polymer design\cite{polymerdesign}        & { \url{ reccr.chem.rpi.edu/polymerdesign}} &     RPI                                                                             & ML for polymer design                                                                                                                                                                                                 & materials design tool                                                                                                                          \\ \hline

Matlearn\cite{peterson2021materials} &\url{matlearn.org} & Univ.of Houston & Predict Formation energy and create composition diagrams using ML to guide synthetic chemistry & inorganic materials design tool
\\ \hline
% NOMAD                          & { { nomad-lab.eu/AIToolkit}}                     & Europe                                                                          & AI toolkit code                                                                                                                                                                                                       & difficult to use                                                                                                          \\ \hline
USPEX\cite{glass2006uspex}                          & { \url{ uspex-team.org/en}}           & Skoltech                                                                        & crystal structure prediction                                                                                                                                                                                          & binary program                                                                                       \\ \hline
CALYPSO\cite{wang2015materials}                        & { \url{ calypso.cn/cdg}}                             & Jilin Univ. China                                                                & crystal structure prediction                                                                                                                                                                                          & binary program                                                                                       \\ \hline
JAMIP\cite{zhao2021jamip}  & { \url{ www.jamip-code.com/}}        & Jilin Univ. China & platform for feature engineering, data preprocessing, ML model building,  property calculation, hpc computing management                                                                                              & not web server. tool to run DFT jobs ML                                                                                  \\ \hline
\end{tabular}
\end{table}

\FloatBarrier

The third category of web apps are diverse utility tools for structure and composition analysis including crystal toolkit, phase diagram and others from Materials projects\cite{ceder2010materials}, prototype finder from aflowlib\cite{curtarolo2012aflowlib}, phase diagram tool from OQMD\cite{kirklin2015open}, analysis tools from JARVIS\cite{choudhary2020joint}, Matgenie from USCD \cite{matgenie}, phonon visualizer from MaterialsCloud\cite{talirz2020materials}, and crystal symmetry tool from Bilbao crys-tallographicserver. 

The fourth category of web tools are the materials design tools including polymer designer \cite{polymerdesign}, Matlearn composition explorer \cite{peterson2021materials}, SUNCAT catalysis designer \cite{winther2019catalysis}, and heterostructure designer in JARVIS\cite{choudhary2020joint}.

There are several offline tools that are very useful for materials discovery including the crystal structure prediction softwares such as USPEX\cite{glass2006uspex} and CALYPSO\cite{wang2015materials}. There are also platform tools such as JAMIP which includes property ML models and first principle calculation job managements.

% https://exabyte.io/#news
% https://github.com/carlyman77/MaterialsDiscoveryML
% http://xtalopt.openmolecules.net/xtalcomp/xtalcomp.html

\section{MaterialsAtlas: Platform of Materials Discovery Tools}

The MaterialsAtlas platform includes four types of web apps for supporting exploratory materials discovery including: composition and structure check and validation, materials property prediction, screening of hypothetical materials, and utility tools.

\subsection{Overview of MaterialsAtlas.org Web App Platform}
\label{sec:headings}

% Computational App Boilerplate. Django + Vue.JS + Redis queues + NginX

Materials informatics web apps have special requirements based on their role in the materials discovery pipeline. For characterization tools, most of the time submission and processing of individual materials are enough. However, for screening tasks such as property prediction, usually it is desirable to be able to upload a list of compositions or structures for high-throughput screening. We also need to consider that there may be multiple users using the web servers simultaneously, and we need a job queue system to process those requests one by one. To meet these requirements, we have developed our MaterialsAtlas.org web service using the Django + Vue.JS + Redis queues + NginX technical framework.

\paragraph{System architecture and web app:} MaterialsAtlas uses Django's built in SQLite3 database for storing hypothetical materials found by our generative materials design models \cite{dan2020generative,zhao2021high,song2021computational}. Moreover, a RESTful API framework is used to send data from the Django back-end to the Vue.js front-end and vice versa. For example, a user will input either a chemical formula or element in one of the apps which will then be interpreted through the Django REST framework. The data is then queued as a job using Redis and subsequently, a Python worker is used to input the data into the corresponding app function. Once the worker and job have finished, the result is returned to the front-end to be viewed by the user. MaterialsAtlas also uses Ajax for some of the applications to communicate to our API. On a separate note, Nginx is used as the web application's HTTP server. Additionally, MaterialsAtlas utilizes Nginx to proxy to the back-end and front-end server. For easier deployment, Docker is used to assemble each web-service as containers allowing the web application to work as a whole. 

\paragraph{Backend models}
Python is used as MaterialsAtlas' primary back-end language to compute each application result and write to the Django database.
\paragraph{Job submission}
When integrating a web application with any machine learning model, latency is a large concern. Using Redis' job queue and fast in-memory data storage functionality allows a web application of this nature to run smoothly.

\subsection{Composition and Structure Validation}

\begin{figure}[t!]
  \centering
  \includegraphics[width=\linewidth]{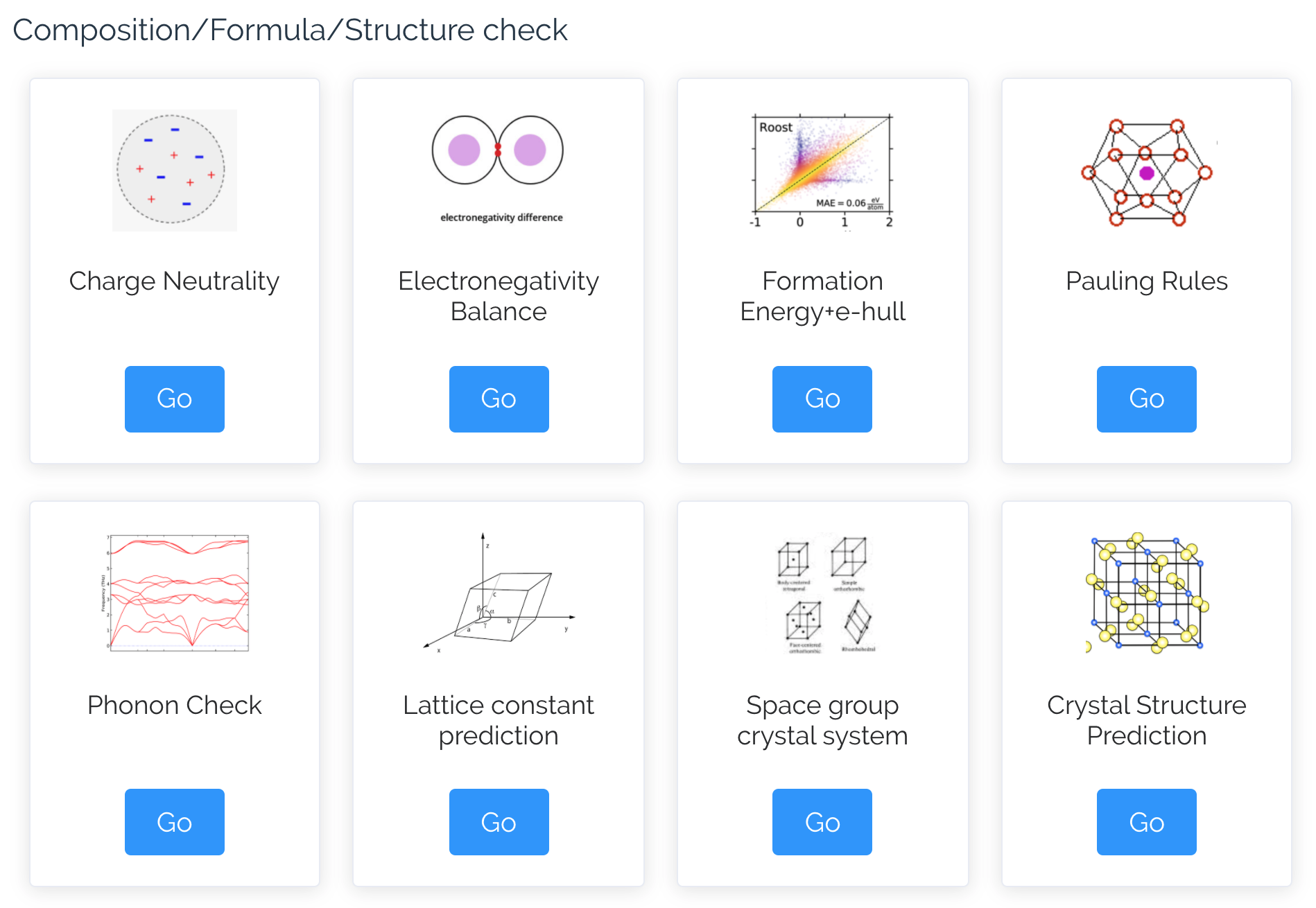}
  \caption{Tools for composition and structure validation and check.}
  \label{fig:validation}
\end{figure}

\paragraph{Chemical validity check:} Given a predicted or generated material composition or structure, there are several steps to verify their physical feasibility. The first quick check of the chemical validity is the charge neutrality and electronegativity balance check (Figure \ref{fig:validation}). These two check algorithms are based on the SMACT package \cite{davies2019smact} with improvements to speed up the enumeration and search process. For both checks, only composition information is needed. Another chemical validation check is the Pauling rules check. Here we only check the input structure against the first three Pauling rules\cite{george2020limited} 
% \hl{It is better to explain those three rules here.}. 

\paragraph{Formation energy and e-above-hull energy check:} Another structure validation step is to check the thermodynamical stability in terms of formation energy calculation. This step is usually done by DFT relaxation and then calculation of their total energy and then formation energy. However, this computation is expensive for large amount of structures. Here, we can first optimize input materials using Bayesian optimization with symmetry relaxation as introduced  by Zuo et al \cite{zuo2021accelerating}. Here, we implemented two ML models for formation energy prediction, one is based on the Roost algorithm \cite{goodall2020predicting} with only the composition as input. This model has demonstrated exceptionally good performance for compound stability prediction among composition only ML models \cite{bartel2020critical}. The other structure based energy prediction model is based on deep global attention graph neural networks \cite{louis2020global} and the input is the crystal structure. The e-above-hull module is forthcoming. 

\paragraph{Phonon calculation: } 
The phonon dispersion relations for the hypothetical materials are important to study the k-space dependence of frequencies of normal modes. It also indicates whether the material is dynamically stable at 0K temperature when there are no imaginary frequencies. We are developing a graph neural network based machine learning model to predict the phonon dispersion relations. This ML-based phonon check module is under development.

\paragraph{Prediction of crystal symmetry (space group and crystal systems) and lattice parameters:} 

Given a materials composition, predicting its structure is very valuable as its many macro-properties such as ion conductivity, thermal conductivity, band gap, and formation energy can be calculated using first principle calculations. However, currently crystal structure prediction is an unsolved problem. In this case, predicting the crystal symmetry such as crystal systems or space groups can be very useful to estimate some of its properties. Here we implement neural network models for space group and crystal system prediction \cite{li2021composition} which have achieved state-of-the-art performance. Another important structure information of crystals is the unit cell parameters, whose precise estimation can greatly help the crystal structure prediction step. Here we implemented a deep neural network model for lattice parameter estimation, which has demonstrated exceptionally good performance for cubic systems and reasonably good results for other crystal systems \cite{li2021mlatticeabc}.

\subsection{Materials property prediction  with composition or structures}
% \hl{steph louis}
% summarize advantage of composition based ML models.
% fast, when data is alot, performance also good...
% cite paper. (roost paper, alphalee, magpie paper).
Depending on the types of features used to train the algorithm, we can categorize the ML properties predictive models as either composition-based or structure-based. Composition-based algorithms, which precede the structure-based algorithms in the literature, have been demonstrated to be reliable, accurate, and even preferred at times \cite{schmidt2019recent}. The composition-based category includes  models that primarily use chemistry-induced descriptors such as elemental representation or chemical composition features \cite{seko2017representation, ward2016general}. Algorithms for these ML models range from very  simple techniques such as Decision-Trees \cite{safavian1991survey} to more complex deep learning algorithms such as Convolutional Neural Networks \cite{o2015introduction} or Graph Neural Networks \cite{zhou2020graph}. 

The adaptation of these composition-based models comes with both advantages and disadvantages. Because these models only use chemical composition descriptors as inputs, their predictive performance heavily relies on the quality of these features; whether they be expert-driven or not. Therefore, the application of these composition-based models requires careful curative steps \cite{schmidt2019recent}. The main disadvantage is that these models omit most of the structural property of the materials. Due to this omission, composition-based models generally offer result in inferior predictive performance compared to structure-based models, especially the size of the dataset is sufficiently large \cite{jha2018elemnet,fung2021benchmarking}. However, thanks to this omission of the materials structure, composite ion based models are more computationally efficient than the structure-based ones. As the composition-based models exclude the materials structures, they don't need to incorporate any methods to extract the structural information \cite{goodall2020predicting}. This omission can be very beneficial in some scenarios since structural-feature extraction is generally very complex and need to be symmetrically invariant \cite{schmidt2019recent}. With just composition descriptors, composition-based models can adapt any simple algorithms such as Decision Trees, Support Vector Machine and still obtain accurate results \cite{schmidt2019recent}. Composition-based models can also adapt more robust algorithms from Deep Learning. Examples of such models include ElemNet (17 layer fully-connected layer)\cite{jha2018elemnet}, Roost (graph neural network)\cite{goodall2020predicting}, and Periodic-table based Convolutional Neural Network \cite{zheng2018machine}. In general, the machine learning adaptations of the compositional-based models tend to be more efficient than their structural-based counter-parts.

% \hl{ summarize structure based property prediction. cite nature benchmarking graph neural networks. Structural descriptors, GNN is much better than descriptors methods, CGCNN, MegNET, DeeperGATGNN..
% sadman}
% \newline
Another category of ML models for materials property prediction is structure based ML models. As almost all materials properties are heavily related to the structures of materials, the structure based ML models for materials property prediction usually achieves greater accuracy than composition based ML models~\cite{xie2018crystal,dunn2020benchmarking}. Structure based models use structure based descriptors~\cite{kajita2017universal,chen2019graph,xie2018crystal}. Structure Graph, Voxel Grids~\cite{zhao2020predicting}, Coulomb Matrix~\cite{rupp2012fast} and Voronoi Tessellation~\cite{chen2020critical} are some of the most popular techniques to represent materials based on the knowledge about their structure. Although models of this category accomplish better prediction results, it can only predict properties of materials whose structures are already known from repositories like Inorganic Crystal Structure Database ($\approx$ 165000 materials)~\cite{bergerhoffinternational} or Materials Project Database ($\approx$ 125,000 materials)~\cite{ceder2010materials} (whereas the cardinality of chemical materials is infinite) and hypothetical materials generated using generative models~\cite{nouira2018crystalgan,dan2020generative}.

Recent studies have shown that when descriptors are learned by deep neural network models, they can predict materials properties with much better accuracy than methods which use descriptors based on physicochemical information~\cite{fung2021benchmarking,louis2020graph}. For doing this, Graph Neural Network (GNN) models have been intensively used as they have shown great success in this task~\cite{xie2018crystal,chen2019graph,schutt2018schnet}. GNN models have a good reputation for finding State-of-the-art (SOTA) performance for various materials property prediction tasks. CGCNN~\cite{xie2018crystal}, MEGNet~\cite{chen2019graph}, GATGNN~\cite{louis2020graph}, SchNet~\cite{schutt2018schnet}, MPNN~\cite{gilmer2017neural} are some of the well known graph neural network models for materials property prediction that use graph representation learning. One of the problems of these existing GNN models is that they cannot go deep, i.e., their performance decreases with increasing number of graph convolution layers as the representation of all the node vectors becomes indistinguishable. This problem is known as the over-smoothing problem~\cite{li2018deeper,chen2020measuring,oono2019graph,louis2020node}, and almost all the GNN models are victim of it. But recently Omee et al. designed a deeper and much improved version of the GATGNN model (DeeperGATGNN) using Differentiable Group Normalization (DGN)~\cite{zhou2020towards} and skip-connections~\cite{he2016deep,jha2021enabling} which allows their DeeperGATGNN to use a high number of graph convolution layers to predict materials property with better accuracy than all the above mentioned GNN models for the five datasets of a recent large-scale benchmark study~\cite{fung2021benchmarking} and the Band Gap dataset from Materials Project Database.

% Practicing deep learning in materials science: An evaluation for predicting the formation energies

\subsection{Property Prediction Tools}
\begin{figure}[ht]
  \centering
  \includegraphics[width=\linewidth]{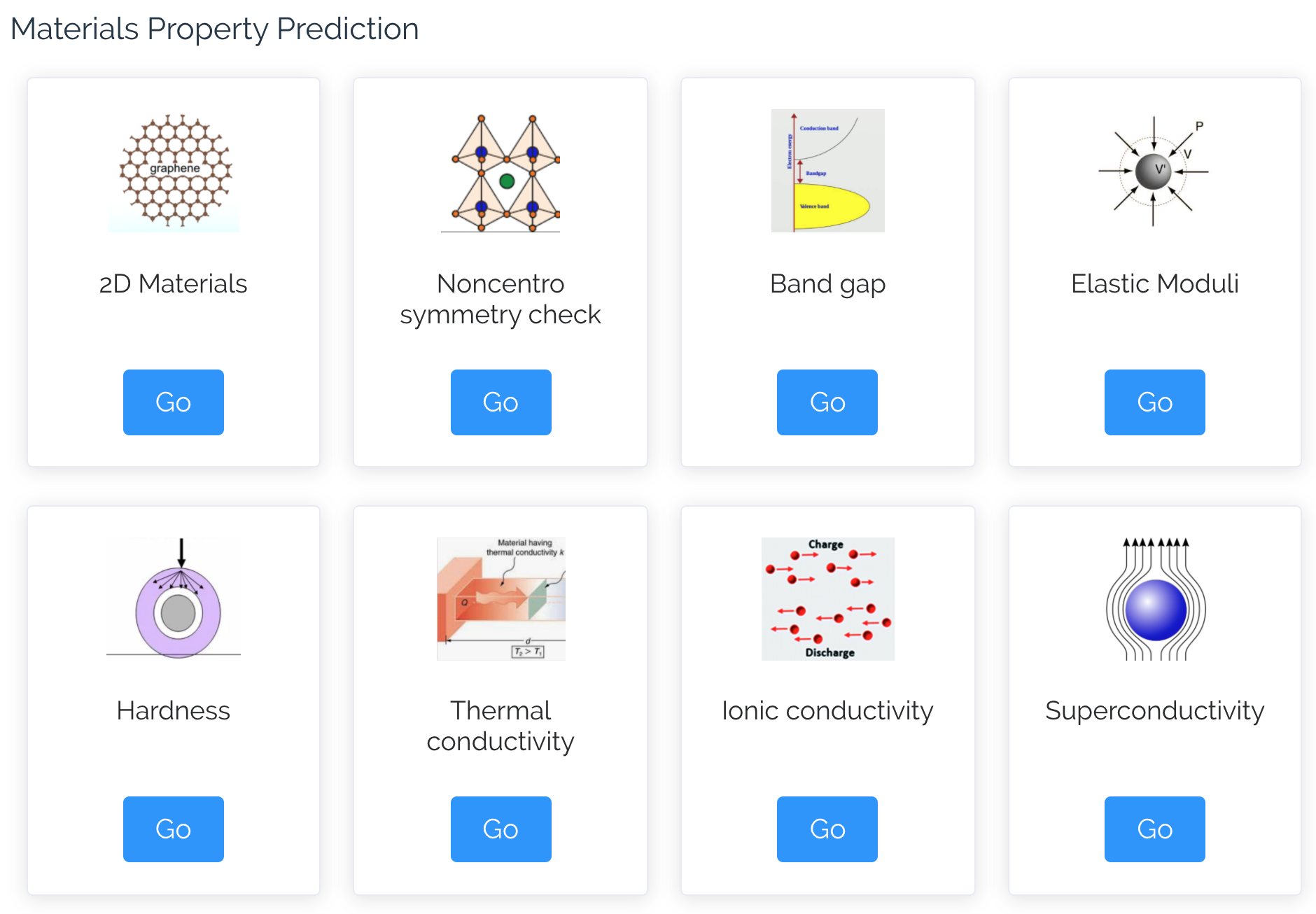}
  \caption{Materials property prediction tools}
  \label{fig:property}
\end{figure}

\begin{table}[h]
\caption{Summary of materials property prediction tools}
\label{table:property tools}
\centering
\begin{tabular}{|c|c|c|c|c|}
\hline
Property prediction  & Model                                                            & Training dataset                                                        & Performance  & Output                                                                                        \\ \hline
2D Materials         & Random Forest                                                    & \begin{tabular}[c]{@{}c@{}}2DMatPedia\\  Material Project\end{tabular} & 88.98\%(Acc) & \begin{tabular}[c]{@{}c@{}}Label\\ probability\end{tabular}                                   \\ \hline
Noncentro Symmetry   & Random Forest                                                    & Material Project                                                        & 84.8\%(Acc)  & \begin{tabular}[c]{@{}c@{}}Label\\  probability\end{tabular}                                  \\ \hline
Band Gap             & \begin{tabular}[c]{@{}c@{}}Roost\\  DeeperGATGNN\end{tabular}  & Material Project                                                        &      0.465(MAE)        & \begin{tabular}[c]{@{}c@{}}Band gap \\ (eV)\end{tabular}                                    \\ \hline
Elastic Moduli       & \begin{tabular}[c]{@{}c@{}}CrabNet\\ DeeperGATGNN \end{tabular}     & \begin{tabular}[c]{@{}c@{}}12858 samples \\ from MP\end{tabular}        &    \makecell{15.7 (MAE, Bulk) \\ 18 (MAE, Shear)\\76.8(MAE, Young's) \\8.7(MAE, Poisson) }          & \begin{tabular}[c]{@{}c@{}}Bulk mod \\ Shear mod \\ Young's mod\\  Poisson ratio\end{tabular} \\ \hline
Hardness             & \begin{tabular}[c]{@{}c@{}}Roost\\ DeeperGATGNN\end{tabular}     & \begin{tabular}[c]{@{}c@{}}12854 samples \\ from MP\end{tabular}        & 0.91($R^2$)     & hardness                                                                               \\ \hline
Thermal Conductivity & \begin{tabular}[c]{@{}c@{}}CrabNet\\ DeeperGATGNN\end{tabular}   & \makecell{2688 samples \\from ICSD  }                                                                  &      5.03(MAE)        & \begin{tabular}[c]{@{}c@{}}thermal conductivity \end{tabular}                        \\ \hline
Ionic Conductivity   & under development                                                &                                                                   N/A      &    N/A          & \begin{tabular}[c]{@{}c@{}}ionic conductivity \\ \end{tabular}                          \\ \hline
Superconductivity    & \begin{tabular}[c]{@{}c@{}}Random Forest\\  CrabNet\end{tabular} & \begin{tabular}[c]{@{}c@{}}25378 samples\\  from supercon \end{tabular}                                                                        &         4.76(MAE)     & transition temperature                                                                        \\ \hline
\end{tabular}                                       
\end{table}

\paragraph{Predicting 2D materials from composition:}
We train a Random Forest classification model to predict whether a given composition forms a 2D or layered structure \cite{song2021computational}. As for the training data, 6,351 2D materials (positive samples) are collect from 2DMatPedia dataset \cite{zhou20192dmatpedia}; 15,959 negative samples are gathered from The Materials Project by removing 2D materials. After training, our model achieves a classification accuracy of 88.98\%. For a given input formula, our model outputs a predicted label (True or False) with corresponding probability in the downloaded results file. Input of multiple formulas are also supported either as a csv file or just typing them into the input box separated by a comma or space. Clicking the 'Check now' button, it will show the found 2D materials; clicking the 'Download results' link, the detailed results will be downloaded.

\paragraph{Predicting noncentrosymmetric materials from composition:} A Random Forest classification model is trained to predict whether a material is noncentrosymmetric \cite{song2020machine}. For training this model, a total of 82,506 samples are collected from the Materials Project by removing those compositions belonging to multiple space groups with conflicting centrosymmetric tendencies; here, 60,687 of them are positive samples and 21,919 are negative samples. The predicted accuracy reaches 84.8\%. The input format and output form are the same as the above method.

\paragraph{Predicting band gap from composition or structure:}
The band gap prediction models are trained with the dataset downloaded from the Materials Project. There are a total of 36,837 samples downloaded. The composition ML model is based on the CrabNet \cite{crabnet}, which uses a transformer self-attention mechanism in the compositionally restricted attention-based network for materials property prediction. Evaluations of over 28 datasets have shown good performance compared to other models. The structure based band gap predictor is based on the dataset downloaded from the Materials Project and trained using the deeperGATGNN graph attention network model \cite{louis2020global}. For a given input formula, this model outputs the predicted band gap values. 

% \hl{sadman, pls. train these 2 models}

\paragraph{Predicting elastic moduli from composition:}
The elastic moduli predicted method with composition is trained by CrabNet as well. Here, we train 4 models to predict bulk mod, shear mod, Young's mod, and poisson ratio.

 \paragraph{Predicting hardness from composition or structure:}
The most recent study uses deep learning for hardness prediction which has shown good performance \cite{mazhnik2020application}. Another study \cite{zhang2021finding} uses 1,062 experimentally measured load-dependent Vickers hardness data extracted from the literature to train the XGBoost ML algorithm using composition-only descriptors with boosting with excellent accuracy (R2 = 0.97). In a related study, XGBoost has been applied to build a temperature dependent Vickers hardness prediction model with R2=0.91 performance using only 593 labelled samples. Here we trained a Roost ML model for composition based hardness prediction.

\paragraph{Predicting thermal conductivity from composition or structure:}
The most recent study on thermal conductivity prediction is from \cite{zhu2021charting} in which graph neural networks (CGCNN) and random forest approaches are combined to build the prediction model. The prediction model is trained with 2,668 ordered and stoichiometric inorganic structures from the ICSD. Here we build a CrabNet \cite{crabnet} model for a composition based prediction model and a deepGATGNN graph neural network model \cite{louis2020global} for structure based predictions. The dataset is downloaded from \cite{gorai2016te}, which contains thermal conductivity values for 2,701 crystal structures contained in the ICSD database. Due to the limited data size, the prediction performance is only for experimental purposes.

% Our composition model is trained with the Crabnet framework\cite{crabnet} and the structure based prediction model is based on our enhanced deeperGATGNN graph neural network \cite{louis2020global}. 

\paragraph{Predicting ion conductivity from composition or structure:} Due to the extremely limited dataset, prediction of ion conductivity, has been very challenging with moderate success by using a set of hand-crafted structural descriptors \cite{sendek2017holistic,sendek2018machine}. This module is under development on our platform.

\paragraph{Predicting superconductor transition temperature from composition:}
We also train a random forest model and a CrabNet model to predict the superconductor transition temperature. The data set is collected from the superCon database \cite{supercon}.

\subsection{Generative Design and Screening for Materials Discovery}
\label{sec:others}

\subsubsection{Deep generative design of materials compositions/formulas}

% \hl{zhaoyong}

Generative models, such as variational autoencoder (VAE)~\cite{kingma2013auto} and Wasserstein generative adversaria network(WGAN)~\cite{arjovsky2017wasserstein}, play an important part in computer vision, audio processing, natural language processing, and molecular science. However, limited works have focused on using generative models to generate virtual inorganic materials (e.g., compositions and crystal structures). There are mainly two directions that researchers use generative models in material science. The first is we use generative models to generate compositions~\cite{dan2020generative,sawada2019study}. Dan et al. propose \cite{dan2020generative} to use WGAN models to generate hypothetical materials compositions which are trained using ICSD dataset. Their models not only can rediscover most compositions from existing materials databases, but also generate many novel compositions that are chemically valid. Here we provide the screening tools for such hypothetical materials.

% \lipsum[10] 
% See Figure \ref{fig:fig1}. Here is how you add footnotes. \footnote{Sample of the first footnote.}
% \lipsum[11] 

\subsubsection{Deep generative design of cubic crystal materials}

% \hl{yong zhao}

Compared to generating virtual materials compositions, generating virtual crystal structures is more helpful for practitioners to find novel materials since many materials properties can only be calculated with the structural information. Several works~\cite{noh2019inverse,court20203,korolev2020machine} based on VAE and~\cite{long2020ccdcgan,kim2020generative,nouira2018crystalgan,zhao2021high} based on GAN have been proposed to generate material structures. CubicGAN proposed by Zhao et al.~\cite{zhao2021high} is the first method that can achieve large scale generative design of novel cubic materials. The authors not only are able to rediscover most of the cubic materials in The Materials Project and ICSD, but also can discover new prototypes with stable materials. In their work\cite{zhao2021high}, they found 31 new prototypes for space groups of Fm$\bar{3}$m, F$\bar{4}$3m, and Pm$\bar{3}$m, of which 4 prototypes contain stable materials. A total of 506 cubic materials have been verified stable by phonon dispersion calculation. Here in our web app platform, we provide the search function for those materials.

% survey of related methods for generative design of structures and introduce cubicgan

% \section{Generative Design and Screening for Materials Discovery}

% \lipsum[2]

\subsubsection{Hypothetical materials screening}

\begin{figure}[ht!]
  \centering
  \includegraphics[width=\linewidth]{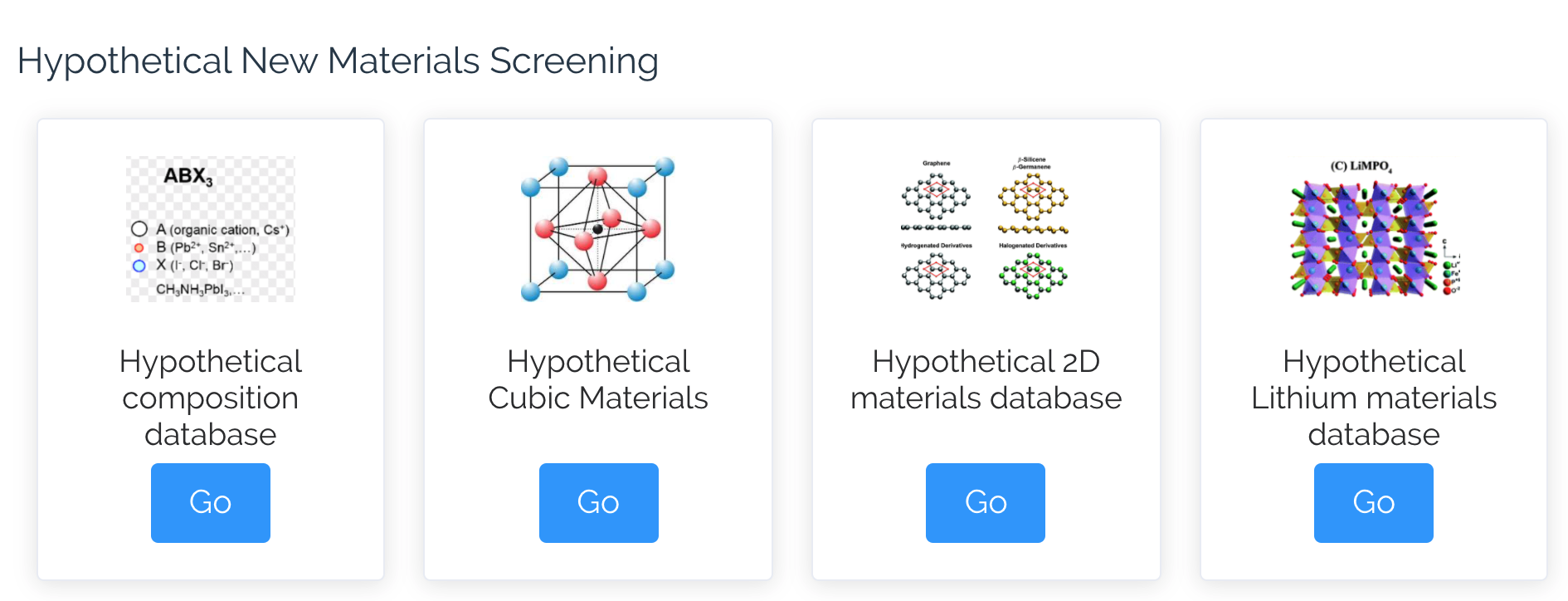}
  \caption{Screening hypothetical materials generated by machine learning or deep learning models.}
  \label{fig:screening}
\end{figure}

One of the major goals for the materials informatics community is to expand the existing materials repositories in terms of materials compositions, structures, and properties, which can help accelerate materials with novel functions. Using our recently developed materials composition generative models (MATGAN) \cite{dan2020generative}, we have generated a large hypothetical materials compositions which are deposited to the databases for screening, hence the Hypothetical composition database (Figure \ref{fig:screening}). For convenience, we also selected those lithium compound candidates and built the Hypothetical lithium materials database. Using our crystal structure generator, the CubicGAN \cite{zhao2021high}, we have created a cubic materials database for screening. Hypothetical materials compositions can also be combined with element substitution based structure prediction to generate new materials database. Finally, we trained a 2D materials classifier which is used to screen the whole hypothetical compositions generated by MATGAN, which are then deposited as Hypothetical 2d materials database.

\FloatBarrier

\subsection{Utility Tools}

\begin{figure}[ht!]
  \centering
  \includegraphics[width=\linewidth]{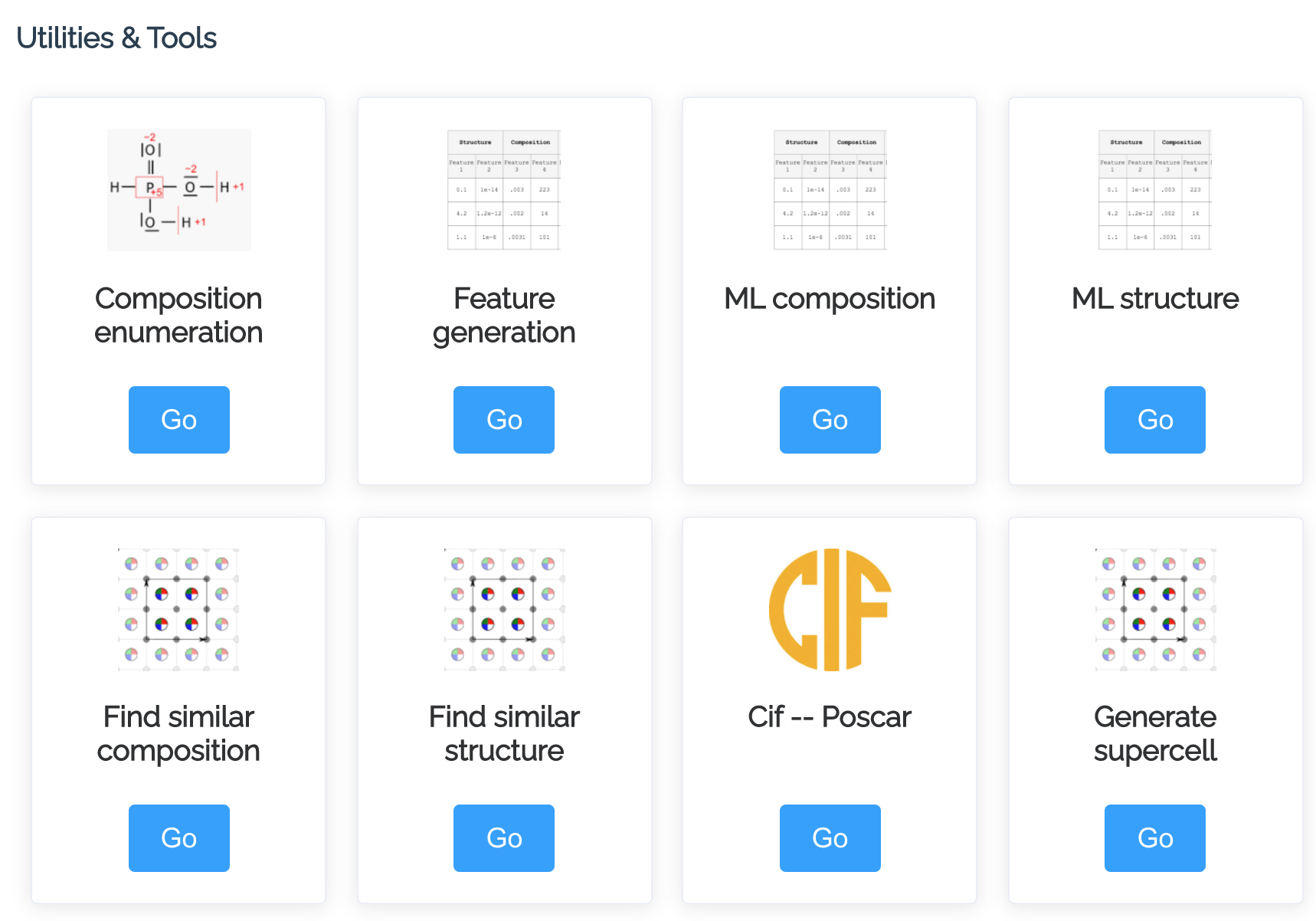}
  \caption{Utility tool web apps.}
  \label{fig:toolbox}
\end{figure}

Several utility tools (Figure\ref{fig:toolbox}) to assist the materials discovery process have been developed and deployed on our platform, including:

\paragraph{Composition enumerator:} Given a number of elements, what are the possible chemically valid formulas that can be synthesized and stable? Due to the oxidation preferences, the number of possibilities are limited and this tool can help the investigator to formulate the target materials composition given a set of elements or an existing formula with one or more dopant elements. With the hypothetical composition, one can then apply crystal structure prediction procedures to get its crystal structure and then many properties by deep learning. This composition enumerator is built based on the SMACT materials package \cite{davies2019smact}.

\paragraph{Feature generation:} The very first step for developing materials property prediction models is to generate and select a set of good descriptors. Here we implemented a pipeline that allows users to choose feature combinations from diverse feature types such as composition features, structure features, electronic features, etc. This will greatly simplify the steps for materials scientists without a strong materials informatics background to develop ML models.

\paragraph{Composition based ML models for user-specified property prediction:} We will build a ML pipeline that allows the user to specify the datasets and target property values and the algorithm, the web tool, will then build ML models and report the prediction performance. The input will be a group of materials formulas. 
% \hl{check if we implemented this}

\paragraph{Structure based ML models for user-specified property prediction: } We will build a pipeline that allow the user to train a structure-based deep global attention graph neural network model for property prediction. 

\section{Future work}

In addition to candidate materials composition and structure validation, materials property prediction, and screening of materials, there are several additional tools that can lower the barrier for materials scientists to exploit data-driven materials discovery techniques, which will be added to our platform.

% \subsection{Machine learning pipeline for materials property prediction}

\subsection{Phonon calculation, synthesizablity prediction, crystal structure prediction and more}

% \hl{dilanga} 

One important validation step for newly proposed hypothetical materials is to calculate its mechanical dynamic stability. This can be done by calculating the phonon dispersion spectrum and checking if it contains negative values. However, this phonon calculation is computationally expensive. Here we will plan to build a machine learning based classifier to check if a structure is stable or not. Another module under development is the synthesizability prediction model, which has been shown to be able to achieve good performance for inorganic materials using semi-supervised machine learning models \cite{jang2020structure}. A machine learning module for this function has been developed and will be deployed soon.

Another important function is crystal structure prediction, which is currently mainly done using DFT based global optimization, which is difficult and applicable only to small systems. We are planning to develop and implement deep learning guided crystal structure algorithms inspired by the AlphaFold \cite{jumper2021highly} for protein structure prediction.

\subsection{Extensible servers and API services}

To expand the coverage of functionalities, our MaterialsAtlas web server is open to include third-party web apps for materials research. We welcome any investigator to collaborate with us and deploy their applications on our platform. Only executable code or python code in a Linux environment is needed. Another useful feature is the REST API services so that other web services can call our APIs to do some query or calculation, which has shown great success in Materials Project's Pymatgen APIs.   

\subsection{Visualization and interactive exploration of design space}

\begin{figure}[ht!]
  \centering
  \includegraphics[width=0.6\linewidth]{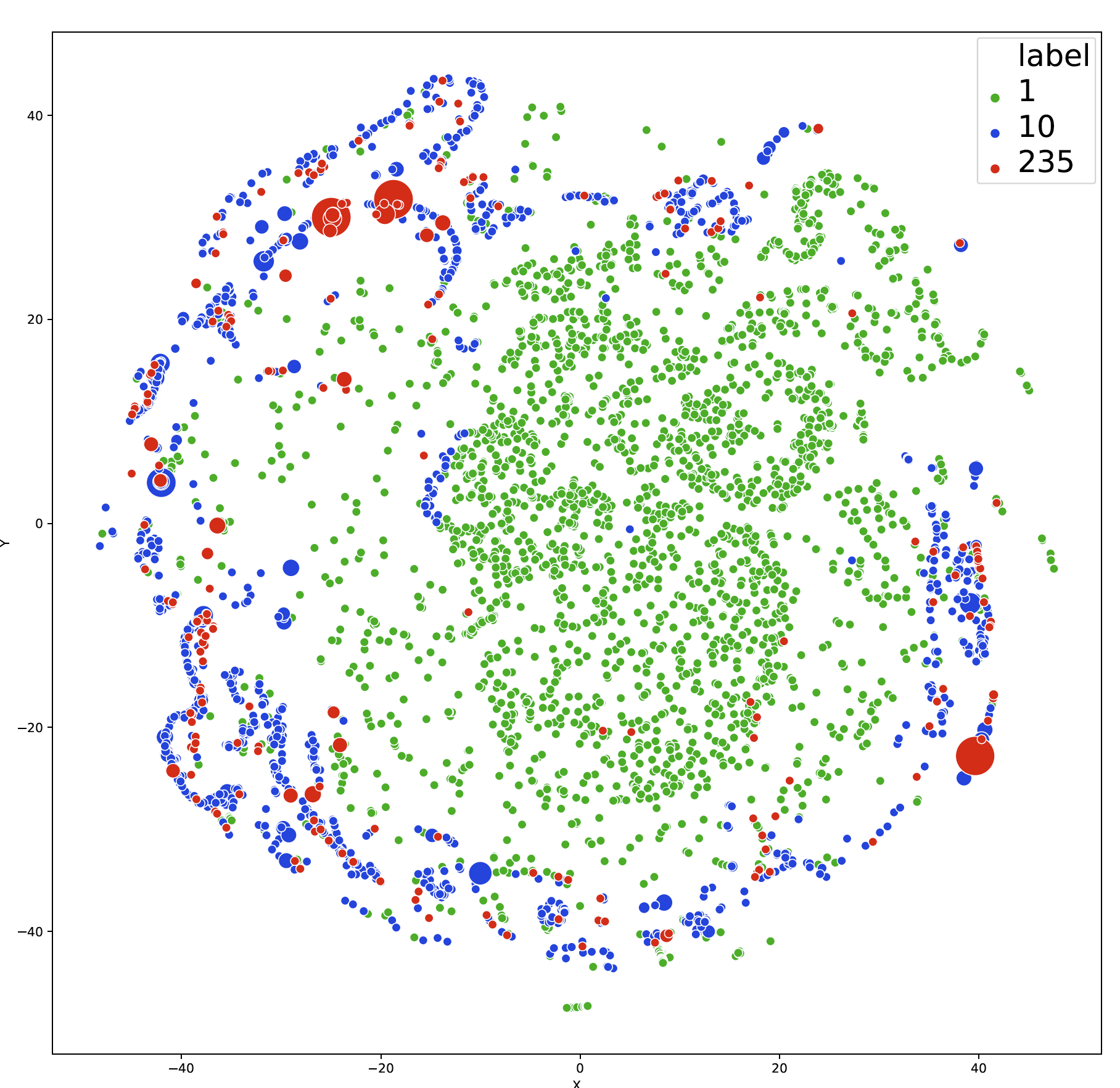}
  \caption{Interactive exploration of thermal conductivity.}
  \label{fig:visual}
\end{figure}

Interactive exploration in the materials design space has big potential to help researchers. We will add modules that support the visualization of materials property distribution among materials in the structural or composition space as shown in Figure\ref{fig:visual}. In this figure, we map the structures into a 2D space using t-sne \cite{van2008visualizing} and XRD representation of the structures. We then annotate those red dots as the samples with annotated thermal conductivity with the dot size representing the magnitude of the thermal conductivity. Such interactive maps will greatly facilitate the search for high performance materials.

\FloatBarrier

\subsection{Utility tools}

\paragraph{Finding similar composition and structures:} In many of the tinkering and explorative studies of the materials design space, it is very helpful to find similar materials and explore their property changes. This search function will help with that. 

Other tools such as structure file conversion and supercell generation will also be added when needed.

\section{Conclusion}
% \lipsum[2]
Despite the rapid progress of machine learning for materials research, a lot of studies have only led to papers with sharing their software while some of them shared their source code but without creating a user-friendly web service or web apps for them. Based on the experience of the bioinformatics field, it is critical for materials informatics researchers to develop and share easy-to-use web apps that wrap their developed algorithms for maximum adoption and usage of such data driven tools in real-life materials discovery and analysis. We have surveyed the status quo of materials informatics web apps and find that they drastically lag behind the bioinformatics community. Here we report our MaterialsAtlas.org web platform that implements and integrates a variety of user-friendly tools for: aiding the materials design space exploration, generation of candidates, and validating the candidates. These tools and those planned together will greatly decrease the barrier for materials researchers without deep computing or machine learning backgrounds to effectively exploit such tools.

\section{Contribution}
Conceptualization, J.H.; methodology,J.H. Y.S.,S.L.,E.S.,Y.Z.; software, J.H., S.S.,Y.S., S.O.; resources, J.H.; writing--original draft preparation, J.H., S.S., Y.S.,S.O.,S.L.,E.S.,Y.Z.; writing--review and editing,  J.H; visualization, J.H. and S.S.; supervision, J.H.;  funding acquisition, J.H.

%%%%%%%%%%%%%%%%%%%%%%%%%%%%%%%%%%%%%%%%%%
\section{Acknowledgement}
The research reported in this work was supported in part by National Science Foundation under the grant and 1940099 and 1905775. The views, perspectives, and content do not necessarily represent the official views of the NSF. We appreciate the help from Xerrak Agha, Daniel Varivoda, Sourin Dey for proofreading.

\bibliographystyle{unsrt}  
\bibliography{references}  %%% Remove comment to use the external .bib file (using bibtex).

\begin{thebibliography}{100}

\bibitem{raccuglia2016machine}
Paul Raccuglia, Katherine~C Elbert, Philip~DF Adler, Casey Falk, Malia~B Wenny,
  Aurelio Mollo, Matthias Zeller, Sorelle~A Friedler, Joshua Schrier, and
  Alexander~J Norquist.
\newblock Machine-learning-assisted materials discovery using failed
  experiments.
\newblock {\em Nature}, 533(7601):73--76, 2016.

\bibitem{ramprasad2017machine}
Rampi Ramprasad, Rohit Batra, Ghanshyam Pilania, Arun Mannodi-Kanakkithodi, and
  Chiho Kim.
\newblock Machine learning in materials informatics: recent applications and
  prospects.
\newblock {\em npj Computational Materials}, 3(1):1--13, 2017.

\bibitem{gubernatis2018machine}
JE~Gubernatis and T~Lookman.
\newblock Machine learning in materials design and discovery: Examples from the
  present and suggestions for the future.
\newblock {\em Physical Review Materials}, 2(12):120301, 2018.

\bibitem{butler2018machine}
Keith~T Butler, Daniel~W Davies, Hugh Cartwright, Olexandr Isayev, and Aron
  Walsh.
\newblock Machine learning for molecular and materials science.
\newblock {\em Nature}, 559(7715):547--555, 2018.

\bibitem{wei2019machine}
Jing Wei, Xuan Chu, Xiang-Yu Sun, Kun Xu, Hui-Xiong Deng, Jigen Chen, Zhongming
  Wei, and Ming Lei.
\newblock Machine learning in materials science.
\newblock {\em InfoMat}, 1(3):338--358, 2019.

\bibitem{morgan2020opportunities}
Dane Morgan and Ryan Jacobs.
\newblock Opportunities and challenges for machine learning in materials
  science.
\newblock {\em Annual Review of Materials Research}, 50:71--103, 2020.

\bibitem{wang2020machine}
Anthony Yu-Tung Wang, Ryan~J Murdock, Steven~K Kauwe, Anton~O Oliynyk,
  Aleksander Gurlo, Jakoah Brgoch, Kristin~A Persson, and Taylor~D Sparks.
\newblock Machine learning for materials scientists: An introductory guide
  toward best practices.
\newblock {\em Chemistry of Materials}, 32(12):4954--4965, 2020.

\bibitem{chen2020machine}
An~Chen, Xu~Zhang, and Zhen Zhou.
\newblock Machine learning: accelerating materials development for energy
  storage and conversion.
\newblock {\em InfoMat}, 2(3):553--576, 2020.

\bibitem{moosavi2020role}
Seyed~Mohamad Moosavi, Kevin~Maik Jablonka, and Berend Smit.
\newblock The role of machine learning in the understanding and design of
  materials.
\newblock {\em Journal of the American Chemical Society}, 142(48):20273--20287,
  2020.

\bibitem{saal2020machine}
James~E Saal, Anton~O Oliynyk, and Bryce Meredig.
\newblock Machine learning in materials discovery: confirmed predictions and
  their underlying approaches.
\newblock {\em Annual Review of Materials Research}, 50:49--69, 2020.

\bibitem{sparks2020machine}
Taylor~D Sparks, Steven~K Kauwe, Marcus~E Parry, Aria~Mansouri Tehrani, and
  Jakoah Brgoch.
\newblock Machine learning for structural materials.
\newblock {\em Annual Review of Materials Research}, 50:27--48, 2020.

\bibitem{chen2020critical}
Chi Chen, Yunxing Zuo, Weike Ye, Xiangguo Li, Zhi Deng, and Shyue~Ping Ong.
\newblock A critical review of machine learning of energy materials.
\newblock {\em Advanced Energy Materials}, 10(8):1903242, 2020.

\bibitem{fehlmann2021aviator}
Tobias Fehlmann, Fabian Kern, Pascal Hirsch, Robin Steinhaus, Dominik Seelow,
  and Andreas Keller.
\newblock Aviator: a web service for monitoring the availability of web
  services.
\newblock {\em Nucleic Acids Research}, 2021.

\bibitem{kern2020lifetime}
Fabian Kern, Tobias Fehlmann, and Andreas Keller.
\newblock On the lifetime of bioinformatics web services.
\newblock {\em Nucleic acids research}, 48(22):12523--12533, 2020.

\bibitem{himanen2019data}
Lauri Himanen, Amber Geurts, Adam~Stuart Foster, and Patrick Rinke.
\newblock Data-driven materials science: status, challenges, and perspectives.
\newblock {\em Advanced Science}, 6(21):1900808, 2019.

\bibitem{ceder2010materials}
G~Ceder and K~Persson.
\newblock The materials project: A materials genome approach, 2010.

\bibitem{curtarolo2012aflowlib}
Stefano Curtarolo, Wahyu Setyawan, Shidong Wang, Junkai Xue, Kesong Yang,
  Richard~H Taylor, Lance~J Nelson, Gus~LW Hart, Stefano Sanvito, Marco
  Buongiorno-Nardelli, et~al.
\newblock Aflowlib. org: A distributed materials properties repository from
  high-throughput ab initio calculations.
\newblock {\em Computational Materials Science}, 58:227--235, 2012.

\bibitem{kirklin2015open}
Scott Kirklin, James~E Saal, Bryce Meredig, Alex Thompson, Jeff~W Doak,
  Muratahan Aykol, Stephan R{\"u}hl, and Chris Wolverton.
\newblock The open quantum materials database (oqmd): assessing the accuracy of
  dft formation energies.
\newblock {\em npj Computational Materials}, 1(1):1--15, 2015.

\bibitem{li2020ai}
Jiali Li, Kaizhuo Lim, Haitao Yang, Zekun Ren, Shreyaa Raghavan, Po-Yen Chen,
  Tonio Buonassisi, and Xiaonan Wang.
\newblock Ai applications through the whole life cycle of material discovery.
\newblock {\em Matter}, 3(2):393--432, 2020.

\bibitem{choudhary2020joint}
Kamal Choudhary, Kevin~F Garrity, Andrew~CE Reid, Brian DeCost, Adam~J Biacchi,
  Angela R~Hight Walker, Zachary Trautt, Jason Hattrick-Simpers, A~Gilad Kusne,
  Andrea Centrone, et~al.
\newblock The joint automated repository for various integrated simulations
  (jarvis) for data-driven materials design.
\newblock {\em npj Computational Materials}, 6(1):1--13, 2020.

\bibitem{dan2020generative}
Yabo Dan, Yong Zhao, Xiang Li, Shaobo Li, Ming Hu, and Jianjun Hu.
\newblock Generative adversarial networks (gan) based efficient sampling of
  chemical composition space for inverse design of inorganic materials.
\newblock {\em npj Computational Materials}, 6(1):1--7, 2020.

\bibitem{zhao2021high}
Yong Zhao, Mohammed Al-Fahdi, Ming Hu, Edirisuriya Siriwardane, Yuqi Song,
  Alireza Nasiri, and Jianjun Hu.
\newblock High-throughput discovery of novel cubic crystal materials using deep
  generative neural networks.
\newblock {\em arXiv preprint arXiv:2102.01880}, 2021.

\bibitem{lu2021computational}
Ziheng Lu.
\newblock Computational discovery of energy materials in the era of big data
  and machine learning: a critical review.
\newblock {\em Materials Reports: Energy}, page 100047, 2021.

\bibitem{stanev2018unsupervised}
Valentin Stanev, Velimir~V Vesselinov, A~Gilad Kusne, Graham Antoszewski,
  Ichiro Takeuchi, and Boian~S Alexandrov.
\newblock Unsupervised phase mapping of x-ray diffraction data by nonnegative
  matrix factorization integrated with custom clustering.
\newblock {\em npj Computational Materials}, 4(1):1--10, 2018.

\bibitem{xiong2017automated}
Zheng Xiong, Yinyan He, Jason~R Hattrick-Simpers, and Jianjun Hu.
\newblock Automated phase segmentation for large-scale x-ray diffraction data
  using a graph-based phase segmentation (gphase) algorithm.
\newblock {\em ACS Combinatorial Science}, 19(3):137--144, 2017.

\bibitem{kaufmann2020crystal}
Kevin Kaufmann, Chaoyi Zhu, Alexander~S Rosengarten, Daniel Maryanovsky,
  Tyler~J Harrington, Eduardo Marin, and Kenneth~S Vecchio.
\newblock Crystal symmetry determination in electron diffraction using machine
  learning.
\newblock {\em Science}, 367(6477):564--568, 2020.

\bibitem{oviedo2019fast}
Felipe Oviedo, Zekun Ren, Shijing Sun, Charles Settens, Zhe Liu, Noor
  Titan~Putri Hartono, Savitha Ramasamy, Brian~L DeCost, Siyu~IP Tian, Giuseppe
  Romano, et~al.
\newblock Fast and interpretable classification of small x-ray diffraction
  datasets using data augmentation and deep neural networks.
\newblock {\em npj Computational Materials}, 5(1):1--9, 2019.

\bibitem{dong2021deep}
Hongyang Dong, Keith~T Butler, Dorota Matras, Stephen~WT Price, Yaroslav
  Odarchenko, Rahul Khatry, Andrew Thompson, Vesna Middelkoop, Simon~DM
  Jacques, Andrew~M Beale, et~al.
\newblock A deep convolutional neural network for real-time full profile
  analysis of big powder diffraction data.
\newblock {\em npj Computational Materials}, 7(1):1--9, 2021.

\bibitem{ozaki2020automated}
Yoshihiko Ozaki, Yuta Suzuki, Takafumi Hawai, Kotaro Saito, Masaki Onishi, and
  Kanta Ono.
\newblock Automated crystal structure analysis based on blackbox optimisation.
\newblock {\em npj Computational Materials}, 6(1):1--7, 2020.

\bibitem{zheng2020random}
Chen Zheng, Chi Chen, Yiming Chen, and Shyue~Ping Ong.
\newblock Random forest models for accurate identification of coordination
  environments from x-ray absorption near-edge structure.
\newblock {\em Patterns}, 1(2):100013, 2020.

\bibitem{crystalsai}
{Crystals.AI}.
\newblock \url{crystals.ai}.
\newblock Accessed: 2021-09-3.

\bibitem{gaultois2016perspective}
Michael~W Gaultois, Anton~O Oliynyk, Arthur Mar, Taylor~D Sparks, Gregory~J
  Mulholland, and Bryce Meredig.
\newblock Perspective: Web-based machine learning models for real-time
  screening of thermoelectric materials properties.
\newblock {\em Apl Materials}, 4(5):053213, 2016.

\bibitem{tanifuji2019materials}
Mikiko Tanifuji, Asahiko Matsuda, and Hideki Yoshikawa.
\newblock Materials data platform-a fair system for data-driven materials
  science.
\newblock In {\em 2019 8th International Congress on Advanced Applied
  Informatics (IIAI-AAI)}, pages 1021--1022. IEEE, 2019.

\bibitem{suncat}
{SUNCAT}.
\newblock \url{catalysis-hub.org}.
\newblock Accessed: 2021-09-3.

\bibitem{peterson2021materials}
Gordon Peterson and Jakoah Brgoch.
\newblock Materials discovery through machine learning formation energy.
\newblock {\em Journal of Physics: Energy}, 2021.

\bibitem{matgenie}
{Matgenie materials analysis web app}.
\newblock \url{http://matgenie.materialsvirtuallab.org/}.
\newblock Accessed: 2021-09-3.

\bibitem{talirz2020materials}
Leopold Talirz, Snehal Kumbhar, Elsa Passaro, Aliaksandr~V Yakutovich, Valeria
  Granata, Fernando Gargiulo, Marco Borelli, Martin Uhrin, Sebastiaan~P Huber,
  Spyros Zoupanos, et~al.
\newblock Materials cloud, a platform for open computational science.
\newblock {\em Scientific data}, 7(1):1--12, 2020.

\bibitem{aroyo2006bilbao}
Mois~Ilia Aroyo, Juan~Manuel Perez-Mato, Cesar Capillas, Eli Kroumova,
  Svetoslav Ivantchev, Gotzon Madariaga, Asen Kirov, and Hans Wondratschek.
\newblock Bilbao crystallographic server: I. databases and crystallographic
  computing programs.
\newblock {\em Zeitschrift f{\"u}r Kristallographie-Crystalline Materials},
  221(1):15--27, 2006.

\bibitem{winther2019catalysis}
Kirsten~T Winther, Max~J Hoffmann, Jacob~R Boes, Osman Mamun, Michal Bajdich,
  and Thomas Bligaard.
\newblock Catalysis-hub. org, an open electronic structure database for surface
  reactions.
\newblock {\em Scientific data}, 6(1):1--10, 2019.

\bibitem{polymerdesign}
{Polymer design}.
\newblock \url{reccr.chem.rpi.edu/polymerdesign}.
\newblock Accessed: 2021-09-3.

\bibitem{glass2006uspex}
Colin~W Glass, Artem~R Oganov, and Nikolaus Hansen.
\newblock Uspex—evolutionary crystal structure prediction.
\newblock {\em Computer physics communications}, 175(11-12):713--720, 2006.

\bibitem{wang2015materials}
Yanchao Wang, Jian Lv, Li~Zhu, Shaohua Lu, Ketao Yin, Quan Li, Hui Wang, Lijun
  Zhang, and Yanming Ma.
\newblock Materials discovery via calypso methodology.
\newblock {\em Journal of Physics: Condensed Matter}, 27(20):203203, 2015.

\bibitem{zhao2021jamip}
Xin-Gang Zhao, Kun Zhou, Bangyu Xing, Ruoting Zhao, Shulin Luo, Tianshu Li,
  Yuanhui Sun, Guangren Na, Jiahao Xie, Xiaoyu Yang, et~al.
\newblock Jamip: an artificial-intelligence aided data-driven infrastructure
  for computational materials informatics.
\newblock {\em Science Bulletin}, 2021.

\bibitem{song2021computational}
Yuqi Song, Edirisuriya M~Dilanga Siriwardane, Yong Zhao, and Jianjun Hu.
\newblock Computational discovery of new 2d materials using deep learning
  generative models.
\newblock {\em ACS Applied Materials \& Interfaces}, 2021.

\bibitem{davies2019smact}
Daniel~W Davies, Keith~T Butler, Adam~J Jackson, Jonathan~M Skelton, Kazuki
  Morita, and Aron Walsh.
\newblock Smact: Semiconducting materials by analogy and chemical theory.
\newblock {\em Journal of Open Source Software}, 4(38):1361, 2019.

\bibitem{george2020limited}
Janine George, David Waroquiers, Davide Di~Stefano, Guido Petretto, Gian-Marco
  Rignanese, and Geoffroy Hautier.
\newblock The limited predictive power of the pauling rules.
\newblock {\em Angewandte Chemie}, 132(19):7639--7645, 2020.

\bibitem{zuo2021accelerating}
Yunxing Zuo, Mingde Qin, Chi Chen, Weike Ye, Xiangguo Li, Jian Luo, and
  Shyue~Ping Ong.
\newblock Accelerating materials discovery with bayesian optimization and graph
  deep learning, 2021.

\bibitem{goodall2020predicting}
Rhys~EA Goodall and Alpha~A Lee.
\newblock Predicting materials properties without crystal structure: Deep
  representation learning from stoichiometry.
\newblock {\em Nature communications}, 11(1):1--9, 2020.

\bibitem{bartel2020critical}
Christopher~J Bartel, Amalie Trewartha, Qi~Wang, Alexander Dunn, Anubhav Jain,
  and Gerbrand Ceder.
\newblock A critical examination of compound stability predictions from
  machine-learned formation energies.
\newblock {\em npj Computational Materials}, 6(1):1--11, 2020.

\bibitem{louis2020global}
Steph-Yves Louis, Yong Zhao, Alireza Nasiri, Xiran Wong, Yuqi Song, Fei Liu,
  and Jianjun Hu.
\newblock Global attention based graph convolutional neural networks for
  improved materials property prediction.
\newblock {\em arXiv preprint arXiv:2003.13379}, 2020.

\bibitem{li2021composition}
Yuxin Li, Rongzhi Dong, Wenhui Yang, and Jianjun Hu.
\newblock Composition based crystal materials symmetry prediction using machine
  learning with enhanced descriptors.
\newblock {\em arXiv preprint arXiv:2105.07303}, 2021.

\bibitem{li2021mlatticeabc}
Yuxin Li, Wenhui Yang, Rongzhi Dong, and Jianjun Hu.
\newblock Mlatticeabc: generic lattice constant prediction of crystal materials
  using machine learning.
\newblock {\em ACS omega}, 6(17):11585--11594, 2021.

\bibitem{schmidt2019recent}
Jonathan Schmidt, M{\'a}rio~RG Marques, Silvana Botti, and Miguel~AL Marques.
\newblock Recent advances and applications of machine learning in solid-state
  materials science.
\newblock {\em npj Computational Materials}, 5(1):1--36, 2019.

\bibitem{seko2017representation}
Atsuto Seko, Hiroyuki Hayashi, Keita Nakayama, Akira Takahashi, and Isao
  Tanaka.
\newblock Representation of compounds for machine-learning prediction of
  physical properties.
\newblock {\em Physical Review B}, 95(14):144110, 2017.

\bibitem{ward2016general}
Logan Ward, Ankit Agrawal, Alok Choudhary, and Christopher Wolverton.
\newblock A general-purpose machine learning framework for predicting
  properties of inorganic materials.
\newblock {\em npj Computational Materials}, 2(1):1--7, 2016.

\bibitem{safavian1991survey}
S~Rasoul Safavian and David Landgrebe.
\newblock A survey of decision tree classifier methodology.
\newblock {\em IEEE transactions on systems, man, and cybernetics},
  21(3):660--674, 1991.

\bibitem{o2015introduction}
Keiron O'Shea and Ryan Nash.
\newblock An introduction to convolutional neural networks.
\newblock {\em arXiv preprint arXiv:1511.08458}, 2015.

\bibitem{zhou2020graph}
Jie Zhou, Ganqu Cui, Shengding Hu, Zhengyan Zhang, Cheng Yang, Zhiyuan Liu,
  Lifeng Wang, Changcheng Li, and Maosong Sun.
\newblock Graph neural networks: A review of methods and applications.
\newblock {\em AI Open}, 1:57--81, 2020.

\bibitem{jha2018elemnet}
Dipendra Jha, Logan Ward, Arindam Paul, Wei-keng Liao, Alok Choudhary, Chris
  Wolverton, and Ankit Agrawal.
\newblock Elemnet: Deep learning the chemistry of materials from only elemental
  composition.
\newblock {\em Scientific reports}, 8(1):1--13, 2018.

\bibitem{fung2021benchmarking}
Victor Fung, Jiaxin Zhang, Eric Juarez, and Bobby~G Sumpter.
\newblock Benchmarking graph neural networks for materials chemistry.
\newblock {\em npj Computational Materials}, 7(1):1--8, 2021.

\bibitem{zheng2018machine}
Xiaolong Zheng, Peng Zheng, and Rui-Zhi Zhang.
\newblock Machine learning material properties from the periodic table using
  convolutional neural networks.
\newblock {\em Chemical science}, 9(44):8426--8432, 2018.

\bibitem{xie2018crystal}
Tian Xie and Jeffrey~C Grossman.
\newblock Crystal graph convolutional neural networks for an accurate and
  interpretable prediction of material properties.
\newblock {\em Physical review letters}, 120(14):145301, 2018.

\bibitem{dunn2020benchmarking}
Alexander Dunn, Qi~Wang, Alex Ganose, Daniel Dopp, and Anubhav Jain.
\newblock Benchmarking materials property prediction methods: the matbench test
  set and automatminer reference algorithm.
\newblock {\em npj Computational Materials}, 6(1):1--10, 2020.

\bibitem{kajita2017universal}
Seiji Kajita, Nobuko Ohba, Ryosuke Jinnouchi, and Ryoji Asahi.
\newblock A universal 3d voxel descriptor for solid-state material informatics
  with deep convolutional neural networks.
\newblock {\em Scientific reports}, 7(1):1--9, 2017.

\bibitem{chen2019graph}
Chi Chen, Weike Ye, Yunxing Zuo, Chen Zheng, and Shyue~Ping Ong.
\newblock Graph networks as a universal machine learning framework for
  molecules and crystals.
\newblock {\em Chemistry of Materials}, 31(9):3564--3572, 2019.

\bibitem{zhao2020predicting}
Yong Zhao, Kunpeng Yuan, Yinqiao Liu, Steph-Yves Louis, Ming Hu, and Jianjun
  Hu.
\newblock Predicting elastic properties of materials from electronic charge
  density using 3d deep convolutional neural networks.
\newblock {\em The Journal of Physical Chemistry C}, 124(31):17262--17273,
  2020.

\bibitem{rupp2012fast}
Matthias Rupp, Alexandre Tkatchenko, Klaus-Robert M{\"u}ller, and O~Anatole
  Von~Lilienfeld.
\newblock Fast and accurate modeling of molecular atomization energies with
  machine learning.
\newblock {\em Physical review letters}, 108(5):058301, 2012.

\bibitem{bergerhoffinternational}
G~Bergerhoff and R~Sievers.
\newblock International union of crystallography: Chester, uk, 1987.

\bibitem{nouira2018crystalgan}
Asma Nouira, Nataliya Sokolovska, and Jean-Claude Crivello.
\newblock Crystalgan: learning to discover crystallographic structures with
  generative adversarial networks.
\newblock {\em arXiv preprint arXiv:1810.11203}, 2018.

\bibitem{louis2020graph}
Steph-Yves Louis, Yong Zhao, Alireza Nasiri, Xiran Wang, Yuqi Song, Fei Liu,
  and Jianjun Hu.
\newblock Graph convolutional neural networks with global attention for
  improved materials property prediction.
\newblock {\em Physical Chemistry Chemical Physics}, 22(32):18141--18148, 2020.

\bibitem{schutt2018schnet}
Kristof~T Sch{\"u}tt, Huziel~E Sauceda, P-J Kindermans, Alexandre Tkatchenko,
  and K-R M{\"u}ller.
\newblock Schnet--a deep learning architecture for molecules and materials.
\newblock {\em The Journal of Chemical Physics}, 148(24):241722, 2018.

\bibitem{gilmer2017neural}
Justin Gilmer, Samuel~S Schoenholz, Patrick~F Riley, Oriol Vinyals, and
  George~E Dahl.
\newblock Neural message passing for quantum chemistry.
\newblock In {\em International conference on machine learning}, pages
  1263--1272. PMLR, 2017.

\bibitem{li2018deeper}
Qimai Li, Zhichao Han, and Xiao-Ming Wu.
\newblock Deeper insights into graph convolutional networks for semi-supervised
  learning.
\newblock In {\em Thirty-Second AAAI conference on artificial intelligence},
  2018.

\bibitem{chen2020measuring}
Deli Chen, Yankai Lin, Wei Li, Peng Li, Jie Zhou, and Xu~Sun.
\newblock Measuring and relieving the over-smoothing problem for graph neural
  networks from the topological view.
\newblock In {\em Proceedings of the AAAI Conference on Artificial
  Intelligence}, volume~34, pages 3438--3445, 2020.

\bibitem{oono2019graph}
Kenta Oono and Taiji Suzuki.
\newblock Graph neural networks exponentially lose expressive power for node
  classification.
\newblock {\em arXiv preprint arXiv:1905.10947}, 2019.

\bibitem{louis2020node}
Steph-Yves Louis, Alireza Nasiri, Fatima~Christina Rolland, Cameron Mitro, and
  Jianjun Hu.
\newblock Node-select: A flexible graph neural network based on realistic
  propagation scheme.
\newblock 2020.

\bibitem{zhou2020towards}
Kaixiong Zhou, Xiao Huang, Yuening Li, Daochen Zha, Rui Chen, and Xia Hu.
\newblock Towards deeper graph neural networks with differentiable group
  normalization.
\newblock {\em arXiv preprint arXiv:2006.06972}, 2020.

\bibitem{he2016deep}
Kaiming He, Xiangyu Zhang, Shaoqing Ren, and Jian Sun.
\newblock Deep residual learning for image recognition.
\newblock In {\em Proceedings of the IEEE conference on computer vision and
  pattern recognition}, pages 770--778, 2016.

\bibitem{jha2021enabling}
Dipendra Jha, Vishu Gupta, Logan Ward, Zijiang Yang, Christopher Wolverton, Ian
  Foster, Wei-keng Liao, Alok Choudhary, and Ankit Agrawal.
\newblock Enabling deeper learning on big data for materials informatics
  applications.
\newblock {\em Scientific reports}, 11(1):1--12, 2021.

\bibitem{zhou20192dmatpedia}
Jun Zhou, Lei Shen, Miguel~Dias Costa, Kristin~A Persson, Shyue~Ping Ong,
  Patrick Huck, Yunhao Lu, Xiaoyang Ma, Yiming Chen, Hanmei Tang, et~al.
\newblock 2dmatpedia, an open computational database of two-dimensional
  materials from top-down and bottom-up approaches.
\newblock {\em Scientific data}, 6(1):1--10, 2019.

\bibitem{song2020machine}
Yuqi Song, Joseph Lindsay, Yong Zhao, Alireza Nasiri, Steph-Yves Louis, Jie
  Ling, Ming Hu, and Jianjun Hu.
\newblock Machine learning based prediction of noncentrosymmetric crystal
  materials.
\newblock {\em Computational Materials Science}, 183:109792, 2020.

\bibitem{crabnet}
Anthony Yu-Tung Wang, Steven~K Kauwe, Ryan~J Murdock, and Taylor~D Sparks.
\newblock Compositionally restricted attention-based network for materials
  property predictions.
\newblock {\em npj Computational Materials}, 7(1):1--10, 2021.

\bibitem{mazhnik2020application}
Efim Mazhnik and Artem~R Oganov.
\newblock Application of machine learning methods for predicting new superhard
  materials.
\newblock {\em Journal of Applied Physics}, 128(7):075102, 2020.

\bibitem{zhang2021finding}
Ziyan Zhang, Aria Mansouri~Tehrani, Anton~O Oliynyk, Blake Day, and Jakoah
  Brgoch.
\newblock Finding the next superhard material through ensemble learning.
\newblock {\em Advanced Materials}, 33(5):2005112, 2021.

\bibitem{zhu2021charting}
Taishan Zhu, Ran He, Sheng Gong, Tian Xie, Prashun Gorai, Kornelius Nielsch,
  and Jeffrey~C Grossman.
\newblock Charting lattice thermal conductivity for inorganic crystals and
  discovering rare earth chalcogenides for thermoelectrics.
\newblock {\em Energy \& Environmental Science}, 14(6):3559--3566, 2021.

\bibitem{gorai2016te}
Prashun Gorai, Duanfeng Gao, Brenden Ortiz, Sam Miller, Scott~A Barnett, Thomas
  Mason, Qin Lv, Vladan Stevanovi{\'c}, and Eric~S Toberer.
\newblock Te design lab: A virtual laboratory for thermoelectric material
  design.
\newblock {\em Computational Materials Science}, 112:368--376, 2016.

\bibitem{sendek2017holistic}
Austin~D Sendek, Qian Yang, Ekin~D Cubuk, Karel-Alexander~N Duerloo, Yi~Cui,
  and Evan~J Reed.
\newblock Holistic computational structure screening of more than 12000
  candidates for solid lithium-ion conductor materials.
\newblock {\em Energy \& Environmental Science}, 10(1):306--320, 2017.

\bibitem{sendek2018machine}
Austin~D Sendek, Ekin~D Cubuk, Evan~R Antoniuk, Gowoon Cheon, Yi~Cui, and
  Evan~J Reed.
\newblock Machine learning-assisted discovery of solid li-ion conducting
  materials.
\newblock {\em Chemistry of Materials}, 31(2):342--352, 2018.

\bibitem{supercon}
Ingo Lütkebohle.
\newblock {National Institute of Materials Science, Materials Information
  Station,SuperCon}.
\newblock \url{http://supercon.nims.go.jp/index_en.html}, 2011.
\newblock [Online; accessed 19-July-2021].

\bibitem{kingma2013auto}
Diederik~P Kingma and Max Welling.
\newblock Auto-encoding variational bayes.
\newblock {\em arXiv preprint arXiv:1312.6114}, 2013.

\bibitem{arjovsky2017wasserstein}
Martin Arjovsky, Soumith Chintala, and Léon Bottou.
\newblock Wasserstein gan, 2017.

\bibitem{sawada2019study}
Yoshihide Sawada, Koji Morikawa, and Mikiya Fujii.
\newblock Study of deep generative models for inorganic chemical compositions.
\newblock {\em arXiv preprint arXiv:1910.11499}, 2019.

\bibitem{noh2019inverse}
Juhwan Noh, Jaehoon Kim, Helge~S Stein, Benjamin Sanchez-Lengeling, John~M
  Gregoire, Alan Aspuru-Guzik, and Yousung Jung.
\newblock Inverse design of solid-state materials via a continuous
  representation.
\newblock {\em Matter}, 1(5):1370--1384, 2019.

\bibitem{court20203}
Callum~J Court, Batuhan Yildirim, Apoorv Jain, and Jacqueline~M Cole.
\newblock 3-d inorganic crystal structure generation and property prediction
  via representation learning.
\newblock {\em Journal of chemical information and modeling},
  60(10):4518--4535, 2020.

\bibitem{korolev2020machine}
Vadim Korolev, Artem Mitrofanov, Artem Eliseev, and Valery Tkachenko.
\newblock Machine-learning-assisted search for functional materials over
  extended chemical space.
\newblock {\em Materials Horizons}, 7(10):2710--2718, 2020.

\bibitem{long2020ccdcgan}
Teng Long, Nuno~M Fortunato, Ingo Opahle, Yixuan Zhang, Ilias Samathrakis, Chen
  Shen, Oliver Gutfleisch, and Hongbin Zhang.
\newblock Ccdcgan: Inverse design of crystal structures.
\newblock {\em arXiv preprint arXiv:2007.11228}, 2020.

\bibitem{kim2020generative}
Sungwon Kim, Juhwan Noh, Geun~Ho Gu, Alan Aspuru-Guzik, and Yousung Jung.
\newblock Generative adversarial networks for crystal structure prediction.
\newblock {\em ACS central science}, 6(8):1412--1420, 2020.

\bibitem{jang2020structure}
Jidon Jang, Geun~Ho Gu, Juhwan Noh, Juhwan Kim, and Yousung Jung.
\newblock Structure-based synthesizability prediction of crystals using
  partially supervised learning.
\newblock {\em Journal of the American Chemical Society}, 142(44):18836--18843,
  2020.

\bibitem{jumper2021highly}
John Jumper, Richard Evans, Alexander Pritzel, Tim Green, Michael Figurnov,
  Olaf Ronneberger, Kathryn Tunyasuvunakool, Russ Bates, Augustin
  {\v{Z}}{\'\i}dek, Anna Potapenko, et~al.
\newblock Highly accurate protein structure prediction with alphafold.
\newblock {\em Nature}, page~1, 2021.

\bibitem{van2008visualizing}
Laurens Van~der Maaten and Geoffrey Hinton.
\newblock Visualizing data using t-sne.
\newblock {\em Journal of machine learning research}, 9(11), 2008.

\end{thebibliography}


% George Kour and Raid Saabne.
% \newblock Real-time segmentation of on-line handwritten arabic script.
% \newblock In {\em Frontiers in Handwriting Recognition (ICFHR), 2014 14th
%   International Conference on}, pages 417--422. IEEE, 2014.

% George Kour and Raid Saabne.
% \newblock Fast classification of handwritten on-line arabic characters.
% \newblock In {\em Soft Computing and Pattern Recognition (SoCPaR), 2014 6th
%   International Conference of}, pages 312--318. IEEE, 2014.

% \bibitem{hadash2018estimate}
% Guy Hadash, Einat Kermany, Boaz Carmeli, Ofer Lavi, George Kour, and Alon
% \newblock Estimate and replace: A novel approach to integrating deep neural
%   networks with existing applications.
% \newblock {\em arXiv preprint arXiv:1804.09028}, 2018.

% \end{thebibliography}

\end{document}